\documentclass[twocolumn,aps,prl,showpacs]{revtex4}
\usepackage{epsfig}

\def\be{\begin{equation}}
\def\ee{\end{equation}}
\def\ba{\begin{eqnarray}}
\def\ea{\end{eqnarray}}

\begin{document}
\title{Controlled Generation of Dark Solitons with Phase Imprinting}
\author{Biao Wu, Jie Liu, and Qian Niu}
\affiliation{Department of Physics, The University of Texas,
Austin, Texas 78712-1081}
\date{\today}
\begin{abstract}
\vskip10pt
The generation of dark solitons in Bose-Einstein condensates 
with phase imprinting is studied by mapping
it into the classic problem of a damped driven pendulum. We provide
simple but powerful schemes of designing the phase imprint for various
desired outcomes. We derive a formula for the number of dark solitons 
generated by a given phase step, and also obtain results which 
explain experimental observations.
\end{abstract}

\maketitle

Solitons have been discovered in various classical nonlinear
media, such as fluid, magnetic and optical systems, and have
fascinated physicists for decades for their particle-like 
properties \cite{Luce}. Recently, dark solitons were observed 
in Bose-Einstein condensates (BECs) of dilute atomic
gases \cite{Phillips,Burger}, which are described by a 
macroscopic wave function. Dark solitons are produced by 
engineering the phase of this wave function with a technique known 
as phase imprinting, which was originally
proposed and used to create vortices \cite{Dobrek}.
Phase imprinting is to  shine an off-resonance laser on
a BEC thus create phase steps between 
different parts of the BEC. As a new tool of manipulating a 
wave function, its power and ability have not been studied in a 
systematic manner.

In this Letter we present a thorough analysis of 
the generation of dark solitons with phase imprinting in BECs.
Our study is facilitated by a novel approach, which maps the soliton 
generation problem into a damped driven pendulum problem.
This method makes it easier to find the dark solitons generated
by a given phase step. More importantly, it changes our 
perspective on the problem of soliton generation. With this
method we can show how to design and control the phase steps for 
various desired outcomes, such as
a specific dark soliton and a specified number of solitons.  
We derive a formula, which relates the winding number of
the pendulum motion to the number of dark solitons generated by a 
given phase step. In addition, we study the 
interesting phenomenon that counter-propagating dark solitons may be
generated by one phase step, as observed in Ref.\cite{Burger}, 
and the physics behind it. Although our study is done in the 
context of a BEC, it can easily be applied to fiber optics, 
where dark solitons have potential 
applications in communication \cite{Kivshar}.

We study the phase imprinting on a quasi-one dimensional BEC,
which is realizable experimentally \cite{Mewes} and was indeed 
used to produce dark solitons in BECs \cite{Burger}. On the other
hand, although the magnetic trap is important for the dynamics of BEC dark 
solitons \cite{Zhang}, it has negligible effect on the generation 
of dark solitons, which has a much shorter time scale than
the subsequent dynamics. Therefore, it is sufficient to neglect 
the trap and use the one dimensional nonlinear Schr\"{o}dinger equation
\be\label{eq:nls}
i\frac{\partial \psi(x,t)}{\partial t} = -{1\over 2}
\frac{\partial^2 \psi(x,t)}{\partial x^2} + u_0^2 |\psi(x,t)|^2 \psi(x,t)\,,
\ee
where $x$ is measured in units of $\xi=1\ \mu$m,  a typical length
unit in this type of experiments, $t$ in units of $ m \xi^2\over \hbar$
($m$ is the atomic mass), $\psi$ in units of the square root of $n_0$, 
the average density the condensate, and the speed of sound 
is $u_0=\sqrt{4\pi n_0 a_s \xi^2}$, with $a_s>0$ being the interatomic 
scattering length. For the experiment with rubidium \cite{Burger}, 
we have $u_0\sim 5.5$ and the time of evolution around $10$\,; for sodium 
experiment\cite{Phillips}, we have $u_0\sim 1$ and the time 
of evolution around $30$\,. 

A dark soliton is characterized by a local density minimum 
moving with constant speed against a uniform 
background \cite{Kivshar,Zakharov}. It has three characteristics, the
depth of its density minimum, the phase step over its density notch, 
and its velocity. However, all the three are related to each other 
and can be specified by its velocity.
The nonlinear Schr\"odinger equation is exactly solvable with
the inverse scattering method \cite{Zakharov},  
according to which the generation of dark solitons 
is determined by the Zakharov-Shabat (ZS) eigenvalue equations,
\ba\label{eq:zak1}
\displaystyle i\frac{\partial U_1(x)}{\partial x} + u_0  
\psi(x,0)U_2(x) &=& \lambda U_1(x),\\\nonumber \\
\label{eq:zak2}
\displaystyle i\frac{\partial U_2(x)}{\partial x} - u_0 
\psi^*(x,0)U_1(x) &=& -\lambda U_2(x),
\ea
where $\psi(x,0)$ is the initial condition. The ZS equations can
have discrete eigenvalues $\lambda_i$ with magnitude smaller than $u_0$.
Corresponding to each $\lambda_i$, a dark soliton with 
velocity $-\lambda_i$ is generated.
For phase imprinting, we have $\psi(x,0)=e^{iS(x)}$, where $S(x)$ is 
the imprinted phase. In this letter, for simplicity, we will concentrate 
on the right phase step, which increases monotonically from the 
left to the right, and stays constant at the boundaries,
\be\label{eq:right}
{{\rm d}S(x)\over {\rm d}x}\ge 0\,,\hspace{1.5cm}
{{\rm d}S(x)\over {\rm d}x}=0~~{\rm as}~~~|x|\rightarrow \infty\,.
\ee
It is straightforward to generalize our method  and results to 
the left phase step and more general phase imprinting, 
or even to the density engineering \cite{Carr}.

\begin{figure}[!htb]
\begin{center}
\resizebox *{7.2cm}{3.7cm}{\includegraphics*{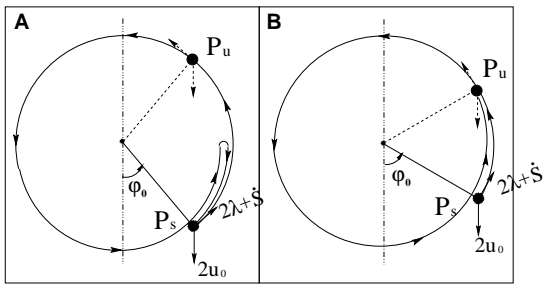}}
\caption{Motions of Pendulum (\ref{eq:pend1}). Trajectories
are schematic, and deliberately distorted when the overlapping occurs.
The vectors are forces.
{\bf A})Motion starting from P$_{\rm s}$ and coming back
to P$_{\rm s}$ after one round of rotation.
{\bf B}) Motion going from the stable fixed point 
P$_{\rm s}$ to the unstable fixed point P$_{\rm u}$.}
\label{fig:pend}
\end{center}
\end{figure}

\begin{figure}[!htb]
\begin{center}
\resizebox *{8.5cm}{6cm}{\includegraphics*{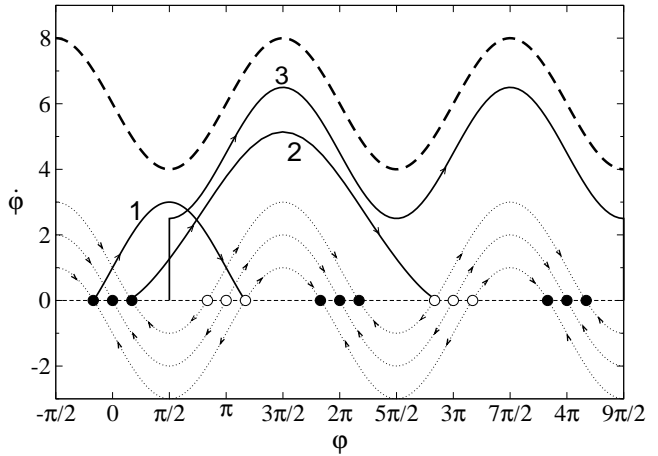}}
\caption{Phase trajectories of a damped driven pendulum of zero mass. 
Dotted line are drawn for constant forces $2\lambda$;
solid lines are for changing forces $2\lambda+\dot{S}$.
The arrows indicate the directions of pendulum motion. 
The black dots are the stable fixed points P$_{\rm s}$; 
the open circles are the unstable fixed point P$_{\rm u}$.}
\label{fig:pp}
\end{center}
\end{figure}

We solve the ZS equations by mapping them into a simple pendulum problem,
which is physically more intuitive and mathematically
much simpler. Note that, for discrete eigenvalues $\lambda_i$,
the ZS wave functions $U_1=\sqrt{\rho_1} e^{i\phi_1}$ and 
$U_2=\sqrt{\rho_2} e^{i\phi_2}$ have the boundary conditions
\be\label{eq:bb1}
\rho_1,\rho_2\rightarrow 0\,,\hspace{0.5cm}
\phi_1,\phi_2\rightarrow {\rm constants}\,,~~{\rm as}~~|x|\rightarrow\infty\,.
\ee 
As is well known, the quantity $|U_1|^2-|U_2|^2$ is conserved and
independent of $x$, which, combined with Eq.(\ref{eq:bb1}), leads
to the conclusion that $|U_1|^2=|U_2|^2=\rho$ for a discrete 
eigenvalue $\lambda_i$.
In light of this, we make the following transformation
\be\label{eq:tran}
U_1=i\sqrt{\rho}\,e^{i(\theta-\varphi+S)/2}\,,\hspace{0.5cm}U_2=
\sqrt{\rho}\,e^{i(\theta+\varphi-S)/2}\,,
\ee
which turns the ZS equations into a pair of very simple
equations 
\ba
\dot{\varphi}=\frac{{\rm d}\varphi}{{\rm d} x}&=&
2\lambda+\dot{S}-2\,u_0\sin\varphi\,,
\label{eq:pend1}\\
\dot{\rho}=\frac{{\rm d}\rho}{{\rm d} x}&=&2\,u_0\,\rho\cos\varphi\,,
\label{eq:pend2}
\ea
with $\dot{\theta}=0$, where the overhead dot denotes 
the spatial derivative. Remarkably, equation (\ref{eq:pend1})
involves only $\varphi$, and can be viewed as a damped massless 
pendulum driven by the force $2\lambda+\dot{S}$ if we regard 
$x$ as time. This pendulum has two fixed points,
P$_{\rm s}$ and P$_{\rm u}$, when $\dot{S}=0$ and $|\lambda|< u_0$.
The point P$_{\rm s}$ is at $\varphi_0=\sin^{-1}{\lambda\over u_0}$  
and is stable; the other one P$_{\rm u}$ is at $\pi-\varphi_0$ and is 
unstable, as shown in Fig\,.\ref{fig:pend},\ref{fig:pp}\,. 

The solutions of Eq.(\ref{eq:pend1}) transformed from the ZS 
eigenfunctions for a discrete eigenvalue always start at P$_{\rm s}$ 
and end at P$_{\rm u}$. This can be checked by noticing
that the boundary conditions (\ref{eq:bb1}) become
$\rho\rightarrow 0$ and $\dot{\varphi}\rightarrow 0$, and 
the discrete eigenvalues have magnitude smaller than $u_0$.

The correspondence between the pendulum solutions going from 
P$_{\rm s}$ to P$_{\rm u}$ and dark solitons can be appreciated
in context of the pendulum problem (\ref{eq:pend1}) itself
without referencing to the ZS equations. With a given phase
step, the pendulum equation (\ref{eq:pend1}) has
a solution starting at the stable fixed point P$_{\rm s}$ for
each $|\lambda|< u_0$, as shown in Fig\,.\ref{fig:pend}.
For future convenience, we name this kind of pendulum solutions 
{\it proto-soliton solutions}. Because of the asymptotic behavior, 
$\dot{S}=0$ at $|x|\rightarrow \infty$, the proto-soliton solution
can only end up at  either P$_{\rm s}$  or P$_{\rm u}$. 
Since P$_{\rm s}$ is the stable fixed point, for most values of $\lambda$
the pendulum comes back to P$_{\rm s}$ after several rounds of rotation 
(Fig\,.\ref{fig:pend}A). Only for finite number of $\lambda$'s 
the pendulum will end up at the unstable fixed point 
P$_{\rm u}$ (Fig\,.\ref{fig:pend}B). When this happens, we say the 
proto-soliton solutions become the soliton solutions.
This set of special 
$\lambda$'s are just the discrete eigenvalues $\lambda_i$ of the ZS equations 
while the soliton solutions correspond to the ZS eigenfunctions.

This novel approach has tremendous advantages over the existing 
methods for the study of ZS equations \cite{Jared}. With the above analysis  
it is clear that we can discard Eq.(\ref{eq:pend2}) and focus only on the 
pendulum equation (\ref{eq:pend1}), which is much simpler than the 
ZS equations. As an example, we solve Eq.(\ref{eq:pend1}) for
the case of the ``sudden'' limit in which the phase imprinted is a step 
function, $S(x)=\alpha\Theta(x)$. In this case, the force is a 
$\delta$-function, $\dot{S}=\alpha\,\delta(x)$.
Integrating Eq.(\ref{eq:pend1}), we have
$
\varphi(0_+)-\varphi(0_-)=(\pi-\varphi_0)-\varphi_0=\alpha\,
$
which gives us $\pi-2\varphi_0=\alpha$. So only one soliton is
generated, and its speed is 
$\lambda=u_0\sin\varphi_0=u_0\cos(\alpha /2)$.
This recovers the result in Ref.\cite{Gredeskul}.

More importantly, this method changes our perspective on the problem 
of generating dark solitons. It allows us ask and answer  
the inverse question, ``what phase step is needed to produce 
a specified dark soliton?'' This is achieved with the
following steps:\\
{\bf 1}) pick $\lambda$, the soliton that one wants to create;\\
{\bf 2}) choose a curve, $\dot{\varphi}=f(\varphi)$, connecting
the pair of fixed points P$_{\rm s}$ and P$_{\rm u}$ corresponding
to $\lambda$ in Fig\,.\ref{fig:pp};\\
{\bf 3}) substitute $\dot{\varphi}=f(\varphi)$ into 
 Eq.(\ref{eq:pend1}), solve it for $S(x)$.\\
The obtained phase step $S(x)$ creates the soliton $\lambda$.
Note that any curve lying entirely in the upper half of the plane 
correspond to right phase steps, $\dot{S}>0$.

Obviously, there are infinite number of paths connecting 
a pair of P$_{\rm s}$ and P$_{\rm u}$, thus there are 
infinite number of phase steps that generate a certain dark soliton. 
We seek one best step in terms of expenditure of energy. 
Imprinting a phase step $S(x)$ injects
energy into the system,
\be
E=\int_{-\infty}^{\infty}{\rm d}x {\dot{S}^2\over 2}
=\int_{\varphi_0}^{\pi-\varphi_0}{\rm d}\varphi
{\dot{S}^2\over 4\lambda-4u_0\sin\varphi+2\dot{S}}\,.
\ee
The minimum value  of this energy corresponds to the smallest disturbance to
the system by phase imprinting or also the smallest amount
of noise in the output. Viewing $E$ as a functional of $\dot{S}$, 
using variational analysis, 
we have the phase step of the least disturbance,
\be\label{eq:best}
S(x)=4\,\tan^{-1}\Big({u_0-\lambda\over \sqrt{u_0^2-\lambda^2}}
\tanh(x\sqrt{u_0^2-\lambda^2})\Big)\,.
\ee
This phase step generates a soliton of velocity $-\lambda$, and
possibly other solitons. Curve 1 in Fig\,.\ref{fig:pp}
is one example of this kind.

One interesting aspect of the phase step (\ref{eq:best}) is that
the step height increases as $\lambda$ gets smaller when
$\lambda$ is positive. This agrees with the experimental results 
that the soliton speed decreases when the step height increases
\cite{Phillips,Burger}. More interesting is when $\lambda$ becomes
negative.  For $\lambda<0$, our numerical study shows that
the step (\ref{eq:best}) always generates more than one
solitons with one traveling to the right with speed $|\lambda|$
and the rest to the left. That implies that counter-propagating
solitons can be generated by one phase step, as
observed experimentally \cite{Burger}. This phenomenon is certainly 
not special to the phase step (\ref{eq:best}), it happens to 
any phase step obtained from a path connecting a pair of the fixed points 
of a negative $\lambda$ in the upper half plane of 
Fig\,.\ref{fig:pp}. However, this phenomenon is rather mysterious and 
against physical intuition.

\begin{figure}[!htb]
\begin{center}
\resizebox*{8.5cm}{11.0cm}{\includegraphics*{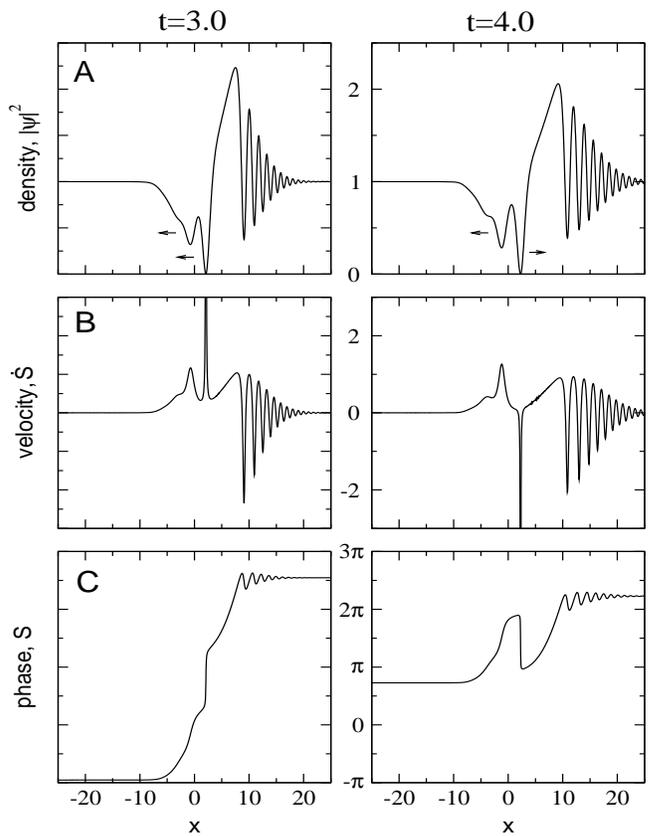}}
\end{center}
\caption{Generation of counter-propagating dark solitons.
Two snapshots are taken at $t=3.0$ and $t=4.0$.
This initial phase step is a tanh function,
$S(x)=\alpha(\tanh(2x/a)+1)/2$\,, with
$u_0=1.0$, $a=3.6$, and $\alpha=3.5\pi$.}
\label{fig:slip}
\end{figure}
Intuitively, once a right phase step (\ref{eq:right})
is imprinted on a BEC cloud, atoms in the imprinted areas will
start moving to the right. Since the atoms outside of the imprinted
area do not move, a dip with a bump to its right will appear 
as a result (see Fig\,.\ref{fig:slip}).
Due to the stronger repulsive interaction from the bump,
the dip will be pushed to the left. As dark solitons come into form
in the dip, one would expect that the dark solitons generated by 
a right phase step would always move to the left. Our
numerical simulation shows that this intuitive picture is only 
correct at the beginning of time evolution. But after a very short time, as
shown in Fig\,.\ref{fig:slip}, one soliton in its early
forming stage may change its direction and start moving to the right. 
More interestingly, 
the phase difference $\Delta S=S(+\infty)-S(-\infty)$ is reduced
by $2\pi$ when this happens, as seen in Fig\,.\ref{fig:slip}C.
This means that the initially imprinted 
the phase difference $\Delta S$ can not always be
maintained in the course of time evolution \cite{field}. 

We are also able to design a phase step which generates a
specified number of solitons. This is based upon the formula
for the number of solitons generated by a given phase step.
To derive the formula,  we need to examine the winding number $W(\lambda)$
of the proto-soliton solution, which
is defined as the number of rounds of rotation that the pendulum
makes in the solution. For a given phase step $S(x)$, 
we have $W(\lambda)\le W(\lambda^{\prime})$
if $\lambda < \lambda^{\prime}$, and 
$W(\lambda^{\prime})-W(\lambda)=1$ when there is only one eigenvalue
$\lambda_i$ between $\lambda$ and 
$\lambda^{\prime}$ \cite{proof}. Therefore, the number of soliton generated
by $S(x)$ is 
$
N_s=W(u_0)-W(-u_0)
$,
and the number of solitons traveling to the right is
$
N_r=W(0)-W(-u_0)\,.
$
As an example we consider a special but still quite general case, a phase
step which can be cut into pieces so that $\dot{S}$ can be considered 
as constant within each piece. Finding the winding number for each piece
and adding them together, we have
\ba
N_s&\approx&{\rm INT}\Big(\int_{-\infty}^{\infty}{{\rm d}x\over \pi}
\sqrt{{\dot{S}^2\over 4}+u_0\dot{S}}\,\Big)\nonumber \\ \label{eq:snum}
&&-{\rm INT}\Big(\int_+{{\rm d}x\over \pi}\sqrt{{\dot{S}^2\over 4}
-u_0\dot{S}}\,\Big)+1\,,
\ea
where the second integral is done over the intervals on which
$\dot{S}>4u_0$,
and INT is the integer function which keeps
only the integer part of a real number.

To imprint a phase which generates exactly $n$ solitons, 
we only need to find a phase step that has $W(u_0)=n$ and $W(-u_0)=0$.
This is achieved by
drawing a path connecting $\varphi=\pi/2$ and $\varphi=2n\pi+\pi/2$, 
the fixed point of the pendulum driven by a constant force $2u_0$. 
At the same time we make
sure it lies under the curve, $\dot{\varphi}=6u_0-2u_0\sin\varphi$, 
the darkened dashed line in Fig\,.\ref{fig:pp}. 
The phase step obtained from this path certainly has $W(u_0)=n$. 
Also this phase step has $W(-u_0)=0$. 
To see this, note that the phase step satisfies
$0\le\dot{S}\le 4u_0$. Therefore,
we always have $\dot{\varphi}\le 0$ at 
$\varphi=\pi/2$ for the case $\lambda=-u_0$ thus 
the pendulum can never pass the position $\pi/2$ and make
a full round of rotation. The simplest example is
curve 3 in Fig\,.\ref{fig:pp}, which represent a linear phase
step generating exactly two solitons.

\begin{figure}[!htb]
\begin{center}
\resizebox*{8.0cm}{6.18cm}{\includegraphics*{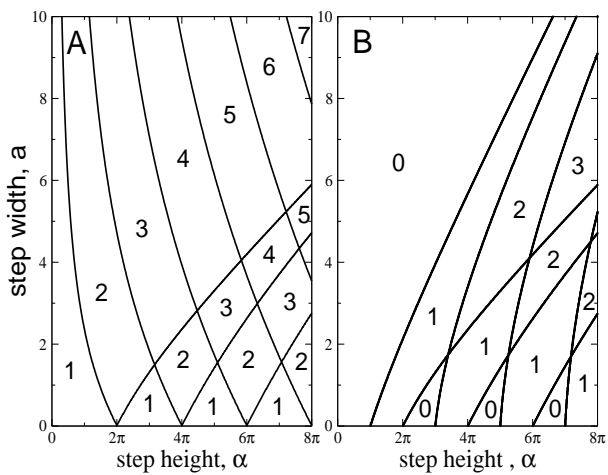}}
\end{center}
\caption{
{\bf A}. Number of Solitons. {\bf B}. Number 
of solitons traveling to the right. }
\label{fig:pd}
\end{figure}
Finally, we apply our results to a simple but very useful
case, the linear phase step:
$\dot{S}=\alpha/a$ for $|x|<a$; 0 for $|x|>\,a/2$.
In the present experiments \cite{Phillips,Burger},
one adjusts the step height and the step width to create
different phase steps, which can be well modelled by the linear 
phase step. For the linear step, Eq.(\ref{eq:snum}) is exact, and it
is used to compute the number of solitons generated for
different step heights $\alpha$ and different step widths $a$, as
shown in Fig\,.\ref{fig:pd}A. 
Also for the simple case, Eq.(\ref{eq:pend1}) can be solved analytically, 
and the eigenvalues $\lambda_i$ are given by
\be\label{eq:linear}
{\lambda^2-u_0^2+{\lambda \alpha\over 2a}\over\sqrt{u_0^2-\lambda^2}} 
={\sqrt{(\lambda+{\alpha\over 2a})^2-u_0^2}\over
\tan(a\sqrt{(\lambda+{\alpha\over 2a})^2-u_0^2}\,)}\,.
\ee
By solving this equation, we obtain the soliton velocities, 
and find the number of solitons traveling to the right, which is
plotted in Fig\,.\ref{fig:pd}B. As a whole, 
Fig\,.\ref{fig:pd} serves as a reference table, where
experimentalists can find the right parameters,
the step height and step width, to generate desired number of solitons.

We thank Roberto Diener for helpful discussions.
This work is supported by the NSF, the Robert A. Welch Foundation, 
and the NSF of China. 
\vskip10pt
\hrule


\begin{thebibliography}{99}
\bibitem{field}According to Ref.\cite{Yang}, the phase difference
$\Delta S$ is defined as the field momentum.  So,
the soliton changes its direction by exchanging momentum 
with the field.
\bibitem{proof} Rigorous proof will be presented elsewhere.
\end{thebibliography}
\end{document}